# Exploring Non-Homogeneity and Dynamicity of High Scale Cloud through Hive and Pig

Kashish Ara Shakil, Mansaf Alam(*Member, IAENG*) and Shuchi Sethi

*Abstract*—Cloud computing deals with heterogeneity and dynamicity at all levels and therefore there is a need to manage resources in such an environment and properly allocate them. Resource planning and scheduling requires a proper understanding of arrival patterns and scheduling of resources. Study of workloads can aid in proper understanding of their associated environment. Google has released its latest version of cluster trace, trace version 2.1 in November 2014.The trace consists of cell information of about 29 days spanning across 700k jobs. This paper deals with statistical analysis of this cluster trace. Since the size of trace is very large, Hive which is a Hadoop distributed file system (HDFS) based platform for querying and analysis of Big data, has been used. Hive was accessed through its Beeswax interface. The data was imported into HDFS through HCatalog. Apart from Hive, Pig which is a scripting language and provides abstraction on top of Hadoop was used. To the best of our knowledge the analytical method adopted by us is novel and has helped in gaining several useful insights. Clustering of jobs and arrival time has been done in this paper using K-means++ clustering followed by analysis of distribution of arrival time of jobs which revealed weibull distribution while resource usage was close to zip-f like distribution and process runtimes revealed heavy tailed distribution.

*Index Terms*—cloud computing, distribution, Google trace K-means++ clustering, Hive, Pig

## I. Introduction

CLOUD computing is gradually being recognized as an emerging paradigm that offers computing resources in a commodity like manner. It promises to provide several advantages and benefits such as provision of availability of resources as a service, dynamic scalability and rapid elasticity [1], [15], [16]. Virtualization is the key to any cloud infrastructure. It helps in achieving scalability and elasticity requirements of cloud [17]-[19]. Cloud computing is nowadays becoming synonymous with non-homogeneity and dynamicity. There occurs non- homogeneity in type of resources available as well as their usage, leading to pitiable system performance and underutilized resources therefore it's important to characterize and study the workloads in cloud so as to properly allocate resources and achieve good performances in such a highly heterogeneous environment. The study of workloads can also help in scheduling of parallel jobs, allocation of processors, resources and load balancing.

Google is an epitome for cloud infrastructure and fairly justifies as an apt cloud environment. In Google trillions of requests are made every day to its datacenters and thus, thousands of jobs are submitted utilizing several resources.

Google has released this information about its workload in form of a trace of workload. Though earlier, trace version 2 was available and several studies have been done in this regard but now Google has released its latest trace version i.e. trace version 2.1[2] in November 2014 and the work carried out in this paper is based on this latest version of the trace. The trace also referred as Clusterdata-2011-2 contains data from more than 11k machines spanning over 700k jobs and represents cell information of about 29 days. Each cell in cluster represents a set of machines of a single cluster. This paper provides a statistical insight into Google cluster trace. It performs an analysis of distribution of arrival times and job based on resource usage. It also performs clustering on the trace data set. In order to carry out the analysis we have used Apache Hive [3] through its beeswax [4] interface along with Apache Pig. To the best of our knowledge the use of such tools in study of workload is novel and no prior analysis has been done using our analytical methods, also no prior study has been done using trace version 2.1.

The major contributions of this paper are:
- Clustering of Jobs based on Resource Utilization
- Clustering of Arrival Time of Jobs
- Analysis of distribution of arrival time of jobs
- Analysis of distribution of jobs based on resource usage
- Analysis of distribution of process runtimes

The work carried out in this paper will aid in further research carried out in a heterogeneous and dynamic environment such as cloud. It will help the researchers in simulating cloud workloads and also in predicting the behavior of applications in cloud. It will also assist in optimizing allocation of resources and management of data in a cloud like environment. Though work has been carried out for management of data in cloud through cloud database management system architecture[5] and k-median clustering [6] but study of a huge workload such as that of Google will further assist in this quest.

The remaining paper has been organized as follows: In

Kashish Ara Shakil is with Department of Computer Science, Jamia Millia Islamia(A Central University), New Delhi, India (+91-9899693456; e-mail: shakilkashish@yahoo.co.in).

Mansaf Alam is with Department of Computer Science, Jamia Millia Islamia(A Central University), New Delhi, India (e-mail: malam2@jmi.ac.in).

Shuchi Sethil is with Department of Computer Science, Jamia Millia Islamia(A Central University), New Delhi, India e-mail: shuchi.sethi@yahoo.com).

section II we study related work which has been done in study of such kind of workload traces. Moving ahead section III gives an insight about the Google cluster trace data set. Section IV deals with statistical analysis of trace dataset. Furthermore Section V shows the clustering of jobs based on K means++ clustering techniques. Section VI gives the conclusion and future directions.

## II. RELATED WORK

Many attempts have been made in order to understand the non-homogeneity and dynamicity of cloud environment. Most of these studies are based on analysis of a large workload such as that of Google. Google cluster is a workload which is a representative of cloud environment. It is highly dynamic and heterogeneous in nature. Several interesting observations can be made by analysis of such large production clusters which can aid further in making scheduling decisions and improving the overall performance of the cluster as a whole.

Apart from study and analysis of Google cluster workload, in [10], [11] analysis of a Map Reduce production cluster has been done. They have analyzed Yahoos Trace data collected from Yahoos M45 supercomputing cluster and has logs of about 10 days. This cluster has around 400 nodes, 4000 processors and approximately 3 terabytes of memory.

They have identified resource utilization patterns, sources of failures of job and job patterns. As per them jobs in cluster followed a long tailed distribution and they have also observed the behavior of users that users run the same job repeatedly and there also exists large error latency in jobs in this cluster.

In [12] analysis of Google trace has been performed, and the authors have concluded that there occurs heterogeneity in all the aspects of the trace i.e. there occurs heterogeneity amongst the resources usage and requirement, as well as heterogeneity in duration of tasks. As per their findings large number of long jobs has stable resource requirements. In our research also we have concluded that there are three types of jobs prevalent in the Google cluster out of which large jobs require more resources that too for a longer duration of time.

In [13] the authors have done a comparative study between grid or other high performance systems such as the ones collected from Grid Workload Archive and Parallel Workload Archive with Google data center. As per them frequency of occurrence of jobs is high in Google trace and the duration of each job is also low as compared to grid systems. Besides these they have also done a study of job priorities i.e. the tasks within the same job have same priority and also a study of job lengths.

In[12] prediction of host workload is done using Bayesian model by capturing features such as predictions and trends of data access and usage. Their results concluded that Bayes method gives high accuracy value with mean error of 0.0014 and provides an improved load prediction accuracy value of 50% as compared to other methods.

In [14] workload classification has been done by identifying workload dimensions and by using k-means algorithm to construct task classes. They have concluded that most of the tasks are short duration ones and most of the resources are consumed by a few tasks.

## III. GOOGLE CLUSTER TRACE OVERVIEW

Google Cluster trace consists of a trace of about 11k machines and 700k jobs running over a span of 29 days. This is the latest version of trace released in November 2014.The cluster consists of racks of machines, each rack consists of several machines packed together. Processing takes place in cluster in form of jobs, where each job is composed of several tasks having varied resource requirements. The trace contains quite a lot of information about machine and job characteristics.

### A. Machine Events

Machine Events information is present in trace in form of timestamp; machine ID of machines, event types of machine such as when machine becomes available, when a machine is removed from cluster and when a machine changes its available resources. Apart from this information such as micro architecture and platform number along with CPU and memory capacity are also available.

### B. Machine Attributes

Machine properties such as speed of clock and external IP address is representative of attributes in the trace.

### C. Job Events

Job Events information is well illustrated in terms of jobs which are running or waiting and scheduling class of jobs indicating latency sensitivity of class.

### D. Task Events

Task Events provides insight into the priority value of tasks such as free priorities, production priorities and monitoring priorities. It also contains information about the request for resources like CPU, RAM and disk usage made by each of the tasks.

### E. Task Constraints

Each task may have many constraints associated with it. The task constraints are represented through timestamp, job id, index value, machine attribute value and name and comparison operators like Less Than, Greater Than, Equal and Not Equal.

### F. Task Resource

The machines in Google cluster make use of Linux containers. The Task resource usage is represented in form of information like start time, end time, job ID, canonical, assigned, unmapped page cache and total page cache memory usage information. Besides these other information's such as disk space usage, I/O time usage, cycle per instruction and memory access per unit are also present in this table.

## IV. STATISTICAL ANALYSIS OF GOOGLE TRACE

In order to carry out statistical analysis of the trace the data was analyzed using Apache Hive and Matlab.Fig.1. shows the workflow diagram of the statistical approach that was adopted by us in our statistical study. We first did initial data filtering using Pig, followed by feature extraction and processing through hive and lastly statistical study was done through Matlab. Apache Hive is software based on data warehouse. It provides facilities for management and analysis of large datasets on distributed storage systems such as Hadoop Distributed File System (HDFS). The use of HDFS ensures scalability and high availability of data at all times. Reading and writing of files to HDFS was conceded out through Apache HCatalog. Querying of data was carried out using HiveQL. Fig. 2. shows a snapshot of query to trace dataset using Hive while Fig.3. shows that of Pig. Pig is a scripting language on top of hadoop that enables analysis of large structured and semi structured data. Furthermore Matlab was used for carrying out other statistical analysis of the trace. Statistical Analysis of the trace dataset was done for the purpose of analyzing the distribution of arrival times of jobs, distribution of the jobs on the basis of resource usage and also clustering of the jobs has been done using K means++ clustering algorithm.

A. *Clustering of Jobs based on Resource Utilization*

In order to find out patterns amongst the jobs in cluster trace K-means++ clustering was performed. The K-means++ algorithm has significant advantages over the traditional k means algorithm which has been used for clustering of workloads in [7], [8].It first finds an initial seeding value which offers considerable advantage by providing faster convergence than the traditional k-means. The K-means++ clustering algorithm defines a method for initializing cluster centers in advance preceded by the standard k –means algorithm. According to K-means++ algorithm [9], First a center is chosen randomly from data points, Then distance between each data point and its nearest center is computed .After this a new data point is chosen at random based on weighted probability distribution. The previous two steps are then repeated until the desired k centers are chosen. After this choice of initial data centers k-means clustering is performed.

Fig. 4. shows the results when clustering was performed for the jobs of the trace using K-means++. From the results obtained three clear clusters of the jobs were visible based on resource usage and thus we could classify the jobs into three categories based on resource usage i.e. we can deduce that the resource usage of jobs in cluster trace is tri-modal in nature. This tri-modal behavior also presented in [8] has been further validated in this paper. Some jobs utilize a large amount of resources, some utilize very small number of resource while others have medium amount of resource

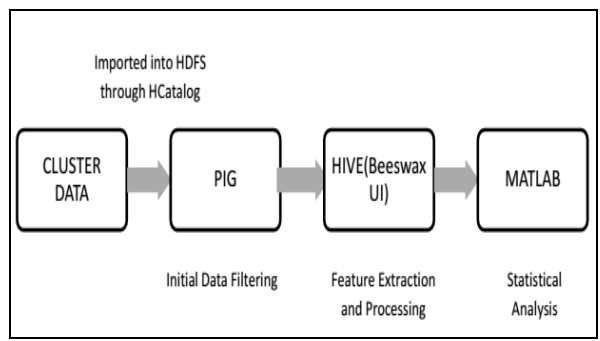

Fig. 1. Workflow diagram of Statistical Analysis

requirements. Thus we can classify the jobs as major resource usage jobs, minor resource usage jobs and mediocre resource usage jobs. For the purpose of experimentation CPU and memory are the resources that have been taken into consideration.

1) Major Resource usage jobs: Major resource usage jobs are the ones that require a lot of resources. These are the jobs that utilize majority of resources in the trace. Since these jobs are the most resource engaging ones therefore such jobs are mostly the longest running jobs in the cluster. Upon analysis of the trace it was revealed that number of such jobs is usually fewer than minor resource usage jobs but they have longer running time and are predominant in the trace and present most of the times. These are jobs that are usually involve complex computations. These jobs can be also be classified as large jobs.

2) Minor resource usage jobs: Minor resource usage jobs are the ones that have minor resource requirements i.e. they require very few resources. Upon analysis of the trace it can be deduced that number of such job types is usually large. Such job types do occur frequently but their running time is usually short.

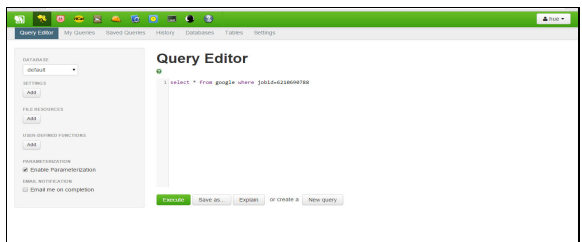

(a)

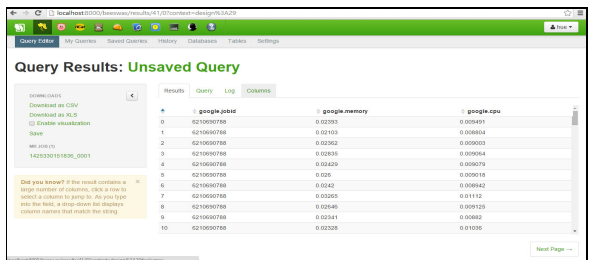

(b)

Fig. 2. (a) Snap Shot of Hives Beeswax UI for processing and querying Trace data (b) Snapshot of output screen after running query on Hive

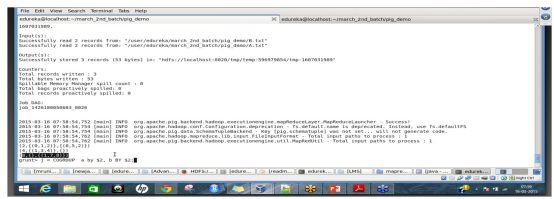

(a)

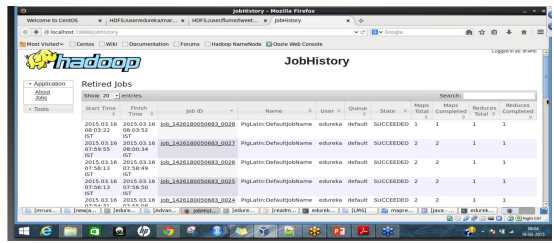

(b)

Fig. 3. (a) Snap Shot of Pig grunt shell (b) Snapshot of output screen showing job history after running a Pig Script

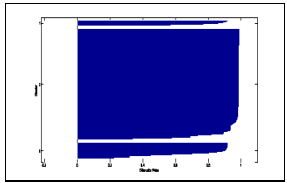
k=3

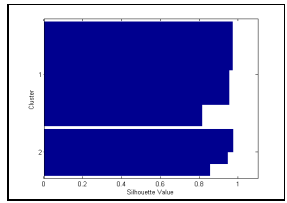
(a) k=2

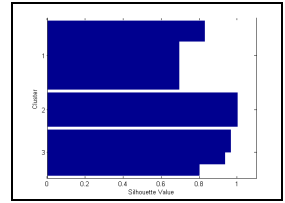
(b) k=3

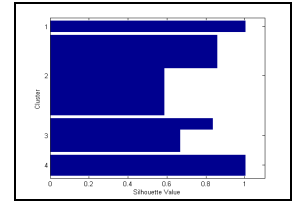
(c) k =4

Fig. 4. K-means++ clustering of jobs

Fig. 5. Clustering of Arrival times of jobs

Such jobs roughly constitute 75 percent of jobs in the trace. These jobs can be also be classified as small jobs.

3) Mediocre resource usage jobs: Mediocre resource usage jobs are the ones whose resource requirements are not as large as the major resource usage jobs and also not as few as the minor resource usage jobs. The number of jobs required by them is in between the two. Their running time is also not very long. These jobs can be also be classified as middle jobs.

### B. Clustering of Arrival Time of Jobs

In order to predict the nature of arrival time of jobs clustering of jobs was done based on the arrival time of jobs using k-means ++ clustering technique, Fig. 5. shows silhouette results after clustering of the jobs based on arrival times. From the figure it's clear that the difference between arrival times of jobs is very less and usually several jobs arrive together in bursts. This can be attributed to the fact that frequency of querying of jobs is usually high.

### C. Study of Distribution of workload parameters

#### 1) Distribution of Arrival Time

At First distribution of arrival times of the jobs was carried out. This distribution was carried out through CDF where CDF refers to cumulative distribution function denoted by F(x). Cumulative density function is defined as the probability that a sample is smaller than or equal to some given value. Equation (1) [21] shows the equation for calculating CDF

$$F(x) = \Pr(X \leq x) \qquad (1)$$

Where Pr(X<x) is probability that sample X is smaller than some value x.

On analysis of Fig. 6. It was inferred that arrival time of jobs in Google trace roughly followed exponential distribution but since the distribution also showed a tail we categorized them under Weibull distribution.

Also it was observed that arrival time of most of the jobs denoted by T($j_i$) is less than 5(equation (2)) i.e.

$$T(j_i) \leq 5 \qquad (2)$$

Weibull distribution is defined by PDF given by equation (3) and CDF given by equation (4)

$$f(x) = \frac{\alpha}{\beta}\left(\frac{x}{\beta}\right)^{\alpha-1} e^{-\left(\frac{x}{\beta}\right)^{\alpha}} \qquad x \geq 0 \qquad (3)$$

Where α, β>0 and are shape and scale parameters respectively.

CDF for Weibull distribution

$$F(x) = 1 - e^{-\left(\frac{x}{\beta}\right)^{\alpha}} \qquad x \geq 0 \qquad (4)$$

From study of the distribution of arrival times we can conclude that the inter arrival time between the jobs is very low.

#### 2) Distribution of jobs based on resource usage

The jobs in Google cluster have already been classified in this paper on the basis of their resource requirements. Further analysis of jobs was then carried out in order to find distribution of jobs with respect to their resource usage.

The resources used for carrying out this analysis included CPU and memory. On analysis of the jobs, which is further illustrated by Fig. 7. on the basis of resource usage it was revealed that the jobs followed a Zipf-like distribution. The PDF for zipf-like distribution is given by equation (5)[21].

$$\Pr(i) \propto \frac{1}{i^{\theta}} \qquad (5)$$

The findings further revealed that some jobs required a lot of resources at all times and such kind of jobs are quite predominant in the cluster and therefore we need effective mechanisms for distribution of such resources and thereby more efficient scheduling algorithms are also required. Furthermore, we can deduce that large jobs usually have high average resource usage as compared to small and middle jobs. If a job is chosen at random from the cluster the probability of it being a large job is high. Thus large jobs are dominant in the cluster.

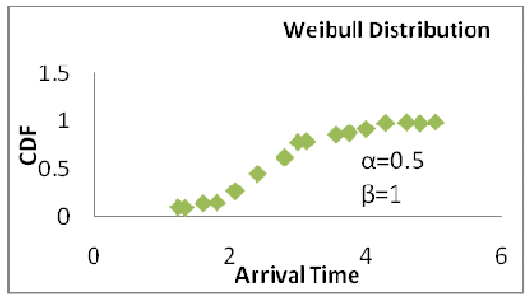

Fig. 6. Weibull distribution CDF for arrival time of jobs

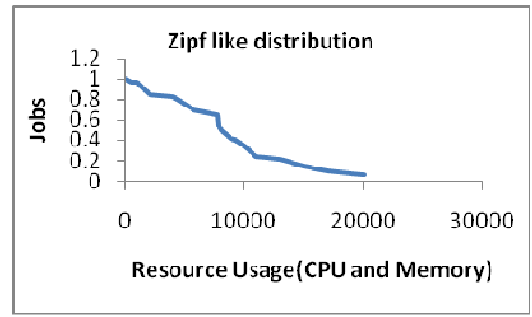

Fig. 7. Zipf like distribution for resource usage

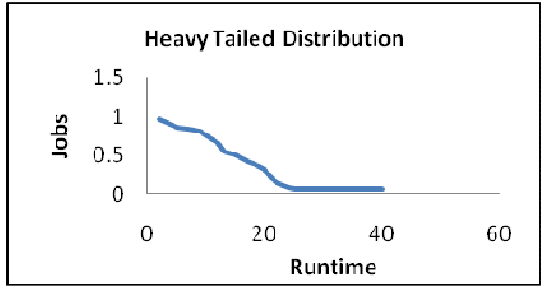

Fig. 8. Heavy tailed distribution for process runtime

### 3) Distribution of process runtimes

On observing the runtimes of jobs, it was revealed that the distribution of runtimes of jobs is skewed with a long tail thus we could conclude that distribution of runtimes of jobs is heavy tailed in cluster trace. According to heavy tailed distribution the probability of occurrence of large values decays. Fig. 8. further shows the distribution of runtimes of jobs in cluster trace. The observation of distribution of runtimes of jobs shows that some jobs had very small values of runtime while others had very large values i.e. there are many small values for runtime of jobs and comparatively fewer large values.

In heavy tailed distribution the tail decays based on power law, equation (6)[21].

$$\overline{F}(x) = \Pr(X > x) x^{-\alpha} \qquad 0 < \alpha \leq 2 \qquad (6)$$

Where, $\overline{F}(x)$ is the survival function i.e. $\overline{F}(x)$ =1-F(x) and α is exponent.

The higher frequency of occurrence of shorter jobs in trace can be attributed to the fact that Google usually caters to smaller set of problems that are less time consuming such a searching for some keywords on the search engine.

## V. CONCLUSION AND FUTURE DIRECTIONS

Cloud computing generally deals with non homogeneous and dynamic environment. Google cluster trace is workload containing cell information of about 29 days spanning across 700k jobs. This paper deals with statistical analysis of Google trace. Google trace contains non homogeneous amount of resources and their usage, study of this trace can help in making useful decisions regarding resource allocation and scheduling. In this paper we have used Hive for analysis of the trace as the size of trace is huge; use of Hive provides the advantage of storage of data in HDFS. In this paper statistical analysis of the trace has been performed, First clustering of jobs based on resource usage has been performed and then clustering of arrival time was done. Apart from this analysis of distribution revealed several interesting results such as arrival time showed weibull distribution and inter arrival time between jobs is also very low. Distribution of jobs based on resource usage showed zipf-like distribution indicating that some jobs required a lot of resources while most of the jobs required fewer amount of resources but their frequency of occurrence is high. Finally distribution of process runtimes revealed heavy tailed distribution.

For future we have planned to expand our study to other cloud workloads such as facebooks map reduce cluster and yahoos M45 cluster logs. We also plan to develop a Google workload cluster simulator.